\documentclass[proof]{WileyASNA-v1}
\usepackage[flushleft]{threeparttable}
\articletype{Article Type}%

\begin{document}

\title{Testing Relativistic Accretion Disk Models with GRO J1655-40}

\author[1,2]{Anastasiya Yilmaz*}

\author[1]{Ji\v{r}í Svoboda}

\author[3]{Victoria Grinberg}
\author[1,4]{Peter G. Boorman}
\author[1]{Michal Bursa}
\author[1]{Michal Dov\v{c}iak}

\authormark{YILMAZ \textsc{et al}}

\address[1]{\orgdiv{Astronomical Institute}, \orgname{Czech Academy of Sciences}, \orgaddress{\state{Prague}, \country{Czech Republic}}}

\address[2]{\orgdiv{Astronomical Institute}, \orgname{Charles University}, \orgaddress{\state{Prague}, \country{Czech Republic}}}

\address[3]{\orgdiv{European Space Research and Technology Centre (ESTEC)}, \orgname{European Space Agency (ESA)}, \orgaddress{\state{Noordwijk}, \country{Netherlarnds}}}

\address[4]{\orgdiv{Cahill Center for Astronomy and Astrophysics}, \orgname{California Institute of Technology}, \orgaddress{\state{Pasadena, CA}, \country{USA}}}

\corres{*Anastasiya Yilmaz, Astronomical Institute of the Czech Academy of Sciences, Bo\v{c}n\'\i -II 1401, Praha 4 Prague, 141~00, Czech Republic. \email{anastasiya.yilmaz@asu.cas.cz}}

\abstract{Black hole X-ray binaries are ideal environments to test the accretion phenomena in the presence of strong gravitational potentials. \texttt{KERRBB} held an important place in the X-ray spectral continuum method for measuring the black hole spin modeling the emission from the innermost regions of the accretion disk. In this work, we present the results of X-ray spectral analysis using publicly available RXTE data of GRO J1655-40 obtained during the 2005 outburst with the two relativistic accretion disk models, \texttt{KERRBB} and \texttt{KYNBB}. Our analysis showed that both models provide identical results with black hole spin measurements, disk temperature, and disk luminosity when the inner edge of the accretion disk is set at the innermost stable circular orbit (ISCO) for the same accretion rates. We couldn't obtain reasonable fits for $\sim$ 89\% of the observations with a fixed black hole spin value at $\mathrm{a_{*}=0.7}$ using both models. Allowing the spin parameter to vary improved the fit statistic significantly with reduced $\rm \chi^{2}$ values being reduced from $\sim$ 10-100 to below 2. Both models revealed black hole spin values varying between $\rm 0.52<a_{*}<0.94$, which can be interpreted as a variable inner edge of the disk throughout different accretion states.}


\jnlcitation{\cname{%
\author{Yilmaz A.}, 
\author{Svoboda J.}, 
\author{Grinberg V.}, 
\author{Boorman P.},
\author{Bursa M.}, and 
\author{Dov\v{c}iak M.}} (\cyear{2022}), 
\ctitle{Testing Relativistic Accretion Disk Models with GRO J1655-40}, \cjournal{Astron.Nachr.}, \cvol{2022}.}

\fundingInfo{This research was supported by the Czech Science Foundation project No. 19-05599Y.}

\maketitle

\section{Introduction}\label{intro}

Black hole X-ray binaries (BHXRBs) are binary systems harboring a stellar-mass black hole as the accretor and a donor star, emitting dominantly in the X-ray band of the electromagnetic spectrum. BHXRBs provide the ideal environment to test the behavior of the matter in strong gravitational potentials around black holes. Most BHXRB systems are known to be transients, spending the big majority of their lives in a quiescent state. As these sources transition through different accretion states throughout the outburst (see, e.g., \citet{Fender04} and \citet{Remillard06}), the contributions from thermal and non-thermal spectral components are observed to evolve which can be traced in a hardness-intensity diagram (HID), transitioning through a "q-shaped" path throughout a full outburst \citep{Fender04, Dunn10} covering the low/hard state (LHS), high/soft state (HSS) and the steep powerlaw or intermediate state. For more on the evolution of these accretion states, see the reviews by \citet{Fender04, Homan05, Remillard06, Done07, Dunn10, Belloni11, Dunn11, Fender12, Fender16}.

GRO J1655-40 was discovered in 1994 \citep{Zhang94} by the Burst and Transient Source Experiment (BATSE) on-board the Compton Gamma Ray Observatory and is one of the most extensively studied Galactic low mass black hole X-ray binaries (LMBXRBs) with a black hole with a mass of ${\rm M_{\mathrm{BH}}}=6.3 \pm 0.5 \: {\rm M_{\odot}}$ and a companion star of ${\rm M_{*}}=2.4 \pm 0.4 \: {\rm M_{\odot}}$ based on optical observations \citep{Orosz97}. Based on optical/NIR observations, however, \citet{Beer02} reported a black hole mass of ${\rm M_{\mathrm{BH}}}=5.4 \pm 0.3 \: {\rm M_{\odot}}$ and a companion star of ${\rm M_{*}}=1.45 \pm 0.35 \: {\rm M_{\odot}}$. The inclination angle has been reported as $\rm i=70.2^{\circ} \pm 1.9^{\circ}$ \citep{Kuulkers00} while \citet{Orosz97, Maccarone02} suggested a difference of $15^{\circ}$ between the binary orbital plane inclination angle and the jet inclination angle based on radio observations. We adopt a distance of d$\rm \sim3.2 \pm 0.2$ kpc based on \citet{Hjellming95}, while \citet{Foellmi06} argued that it should be lower than 1.7 kpc. Different methods to measure the spin on GRO J1655-40 revealed a controversy, with $\rm 0.65< a_{*} <0.75$ obtained from spectral continuum fitting  with the Advanced Satellite for Cosmology and Astrophysics (ASCA) and the Rossi X-Ray Timing Explorer (RXTE) observations \citep{Shafee06} while Fe emission line profile predicts much higher values, with a lower limit on the spin at $\rm a_{*}=0.9$ \citep{Reis09} based on XMM-Newton observations. On the contrary to these measurements pointing towards a high spin case for GRO J1655-40, timing analysis of the entire set of RXTE data using the relativistic precession model \citep{Stella98, Stella99} resulted in a much smaller spin parameter $\rm a_{*} = 0.290 \pm 0.003$ \citep{Motta14}. 

The X-ray continuum fitting has been used extensively to measure the black hole spin and depends on accurately locating the innermost stable circular orbit (ISCO) by modeling the thermal component of the X-ray spectral continuum. In this method, it's a common practice to focus primarily on the HSS to measure the location of $\rm R_{{ISCO}}$ which is determined by the black hole spin and mass as follows 

\begin{equation}\label{r_isco}
	\rm R_{ISCO}=R_{g}\left\{3+A_{2} \pm\left[\left(3-A_{1}\right)\left(3+A_{1}+2 A_{2}\right)\right]^{1 / 2}\right\}
\end{equation}
\begin{equation}\label{A_1}
	\rm A_{1}=1+\left(1-a_{*}^{2}\right)^{1 / 3}\left[\left(1+a_{*}\right)^{1 / 3}+\left(1-a_{*}\right)^{1 / 3}\right]
\end{equation}
\begin{equation}\label{A_2}
	\rm A_{2}= \left(3 a_{*}^{2}+A_{1}^{2}\right)^{1 / 2} 
\end{equation}

\begin{equation}\label{A_2}
	\rm a_{*}=\frac{|\mathbf{J}| c}{G M^{2}} 
\end{equation}

where $\rm J$ is the black hole angular momentum, $\rm G$ is the gravitational constant, $\rm c$ is the speed of light, $\rm M$ is the black hole mass, $\rm R_{g}= G M / c^{2}$ is the gravitational radius and the upper and lower signs describe a prograde disk ($\rm a_{*}>0$) and a retrograde disk ($\rm a_{*}<0$), respectively. 

In this paper, we test the most widely used relativistic disk models \texttt{KERRBB}  and \texttt{KYNBB} by performing an extensive spectral analysis using all of the publicly available RXTE observations during the 2005 outburst of GRO J1655-40. The structure of this paper is as follows: in section \ref{data}, we summarize the RXTE data reduction, provide a review of the relativistic spectral modeling of accretion disks in black hole X-ray binaries and summarize the spectral analysis procedure. In section \ref{results}, we present the results of our analysis and summarize our conclusions in section \ref{conclusions}.

\section{Data Reduction and Analysis}\label{data}

\subsection{RXTE Observations Data Reduction}
RXTE observed GRO J1655-40 extensively throughout the entire mission spanning multiple outbursts between 1996-2011. For the spectral fitting, we used a total of more than 502 observations from the 2005 outburst of GRO J1655-40 \citep{Markwardt05} publicly available on HEASARC (High Energy Astrophysics Science Archive Research Center) archive\footnote{\url{https://heasarc.gsfc.nasa.gov/docs/archive.html}}.

We followed the standard procedure in the RXTE cookbook\footnote{\url{https://heasarc.gsfc.nasa.gov/docs/xte/recipes/cook_book.html}} using \texttt{FTOOLS/HEASOFT} version 6.29\footnote{\url{https://heasarc.gsfc.nasa.gov/docs/software/lheasoft/}} software package for data reduction. We only used the Proportional Counter Unit-2 (PCU-2) of  RXTE’s  Proportional  Counter Array (PCA, \citet{Jahoda06}) with all of its layers included as it was almost always operating throughout the entire RXTE mission and has the best calibration among other PCUs on-board PCA. We removed data lying within 10 minutes of the South Atlantic Anomaly and extracted the source spectra in the \texttt{standard2f} mode which provides the optimal spectral resolution and generated response matrices and background spectra. We then made use of the publicly available tool \texttt{pcacorr} for calibration and applied an additional 0.1\% systematic errors to account for uncertainties in the telescope’s response following \citet{Garcia14}. 

\subsection{Relativistic Spectral Modeling of the Accretion Disks in Black Hole X-ray Binaries}\label{modeling}

Proper spectral modeling of the observed thermal spectrum is crucial to developing a better understanding of the inner structure of the accretion disk and the nature of the accretion flow in the presence of strong gravitational potentials around black holes. It is, therefore, important to include a detailed treatment of relativistic effects on the thermal emission from the innermost region of the accretion disk.

To this date, there have been several attempts to include relativistic accretion disk models in \texttt{XSPEC}. \texttt{KERRBB} \citep{Li05} and \texttt{KYNBB} \citep{Dovciak04} played a significant role in the X-ray spectral continuum method to determine the black hole spin. Both \texttt{KERRBB} and \texttt{KYNBB} assume a \citet{Novikov73} geometrically thin, optically thick accretion disk where orbits around the black hole are Keplerian. Both rely on the ray-tracing technique to compute the spectrum of accretion disks around black holes where the local flux is calculated as a sum of black body radiation with the \citet{Novikov73} radial temperature profile. The space-time around the black hole is described by the Kerr metric and both models are constructed to include relativistic effects on the accretion disk due to strong gravitational fields around black holes (i.e. light bending, self-irradiation or the returning radiation, gravitational redshift, frame dragging and Doppler boost) while computing the observed spectrum. 

Additional to the relativistic effects, \texttt{KERRBB} and \texttt{KYNBB} also account for any deviations of the observed spectrum from the perfect blackbody emission assumed by non-relativistic models like \texttt{DISKBB} and \texttt{DISKPBB} by introducing the color correction factor $\rm f_{col} = T_{col} / T_{eff}$ where $\rm T_{col}$ is the color temperature and $\rm T_{eff}$ is the effective temperature of the accretion disk. This deviation in the observed color temperature becomes prominent when the gas temperature at the surface of the accretion disk is high enough such that the electron scatterings start to dominate over absorptive processes taking place in the disk atmosphere, usually corresponding to photon energies > 1.0 keV \citep{Shakura73}.

\texttt{KYNBB} is a local model that was developed as an extension to the original \texttt{KY} package\footnote{\url{https://projects.asu.cas.cz/stronggravity/kyn}}. While \texttt{KERRBB} describes an accretion disk with an inner edge extending down to the (ISCO), \texttt{KYNBB} offers an option to have the inner edge either as a free parameter or fixed at a certain value other than $\rm R_{ISCO}$. \texttt{KYNBB} also has multiple options one can employ to include the polarization calculations and/or calculate the spectrum only from a selected section of the disk but these additional options are turned off for the entirety of our spectral analysis. For more details about the other model parameters not listed in TABLE \ref{table:model_pars}, see \url{https://projects.asu.cas.cz/stronggravity/kyn}.

\subsection{Spectral Analysis}\label{spectral_analysis}

We used the \texttt{HEASOFT} (v.6.29) package \texttt{XSPEC (v.12.12)} \citep{Arnaud96} and \texttt{PyXspec}\footnote{\url{https://heasarc.gsfc.nasa.gov/xanadu/xspec/python/html/}}, a Python interface to the \texttt{XSPEC} for spectral analysis. To apply $\rm \chi^{2}$ statistics, we grouped spectra to have 25 counts per bin using the \texttt{FTOOLS} task \texttt{grppha} and restricted the energy range to 3-25 keV for all of the PCA spectra. All of the models used for the spectral analysis consist of a thermal disk component (\texttt{KERRBB} or \texttt{KYNBB}) and a simple powerlaw component accounting for the non-thermal component of the continuum (\texttt{POWERLAW} in \texttt{XSPEC}). We use \texttt{TBABS} to account for the ISM absorption. Throughout this paper, we refer to each model (\texttt{TBABS*(KERRBB+POWERLAW))} and \texttt{TBABS*(KYNBB+POWERLAW)}) by the specific disk model used for the analysis. We summarize all of the parameters and their values used in this paper in Table \ref{table:model_pars}. Since both disk models do not have the disk temperature as a model parameter, we evaluated the position of the peak of the disk component in both models in order to obtain the value of the disk temperature. While this peak corresponds to a disk radius not exactly at $\rm R_{ISCO}$ but slightly further out due to the zero-torque boundary condition adopted by \texttt{KERRBB} at the innermost disk radius, this is still a valid approximation of the accretion disk temperature as measured by \texttt{KERRBB}.

\begin{table}[h!]
\centering
	\caption{Model parameter values for \texttt{KERRBB} and \texttt{KYNBB}} 
	\label{table:model_pars}
	\begin{tabular}{p{0.14\textwidth}p{0.14\textwidth}p{0.14\textwidth}}
		\hline \hline
		 Model Parameter & Value & Free/Frozen \\
		\hline
		&\texttt{KERRBB}\\
		\hline

            $\eta$ & 0 & Frozen \\
		Black Hole Spin & -- & Free \\
		Inclination Angle & 70.2 & Frozen \\
		Black Hole Mass & 6.3 & Frozen \\
		Mass Accretion Rate & -- & Free \\
		Distance & 3.2 & Frozen \\
		$\rm f_{col}$ & 1.7 & Frozen\\
            Self-irradiation Flag & 0 & Frozen \\
            Limb-darkening Flag & 0 & Frozen \\
            Normalization & 1 & Frozen\\

		\hline
		&\texttt{KYNBB}\\
		\hline
		Black Hole Spin & -- & Free \\
		Inclination Angle & 70.2 & Frozen \\
		Inner Radius & -- & See Section \ref{spectral_analysis} \\
		Outer Radius & 1000 & Frozen \\
		Black Hole Mass & 6.3 & Frozen \\
		Mass Accretion Rate & -- & Frozen (see Section \ref{spectral_analysis} )\\
		$\rm f_{col}$ & 1.7 & Frozen \\
		Normalization & 9.76 & Frozen\\

		\hline
		
		\hline \hline
		
	\end{tabular}
\end{table}

We used disk fractions to distinguish observations based on how much the disk component is dominating over powerlaw following \citet{Kording06, Dunn08, Dunn10} with the disk fraction defined as \\

\begin{equation}\label{disk_fraction}
 \rm F_{Disk}=\frac{L_{Disk} \;(0.001-100.0 \; keV)}{ L_{Disk} (0.001-100.0 \; keV)+L_{Power} (1.0-100.0 \; keV)}
\end{equation}\\

where $\rm L_{Disk}$ and $\rm L_{Power}$ are the unabsorbed disk luminosity in 0.001-100.0 keV and unabsorbed powerlaw luminosity in 1.0-100.0 keV energy range, respectively. We used the convolution model \texttt{CFLUX} in \texttt{XSPEC} for flux calculations of each component.

We first analyzed the entire set of observations using \texttt{KERRBB} with black hole spin and accretion rate as free parameters. For further analysis, we adopted a selection criteria by assessing reduced $\rm \chi^{2}$ ($\rm \chi^{2}$/d.o.f.) values and excluded the observations with reduced $\rm \chi^{2}>2.0$ to make sure that a simple model consisting of thermal (\texttt{KERRBB} or \texttt{KYNBB}) and non-thermal (\texttt{POWERLAW}) components was able to fit the spectra well to obtain reasonable values for the physical parameters of the system. Then, using the sample after the reduced $\rm \chi^{2}$ selection criteria applied, we set up \texttt{KYNBB} to match \texttt{KERRBB} by fixing the inner disk radius at $\rm R_{ISCO}$ instead of leaving it as a free and set the accretion rate parameter in \texttt{KYNBB}, the photon index and the normalization of \texttt{POWERLAW} in \texttt{TBABS*(KYNBB+POWERLAW)} to the values obtained by \texttt{TBABS*(KERRBB+POWERLAW)}, leaving the spin as the only free parameter remaining in the model for the spectral fitting.

\section{Results}\label{results}
\begin{figure*}
\centering
\includegraphics[width=\linewidth]{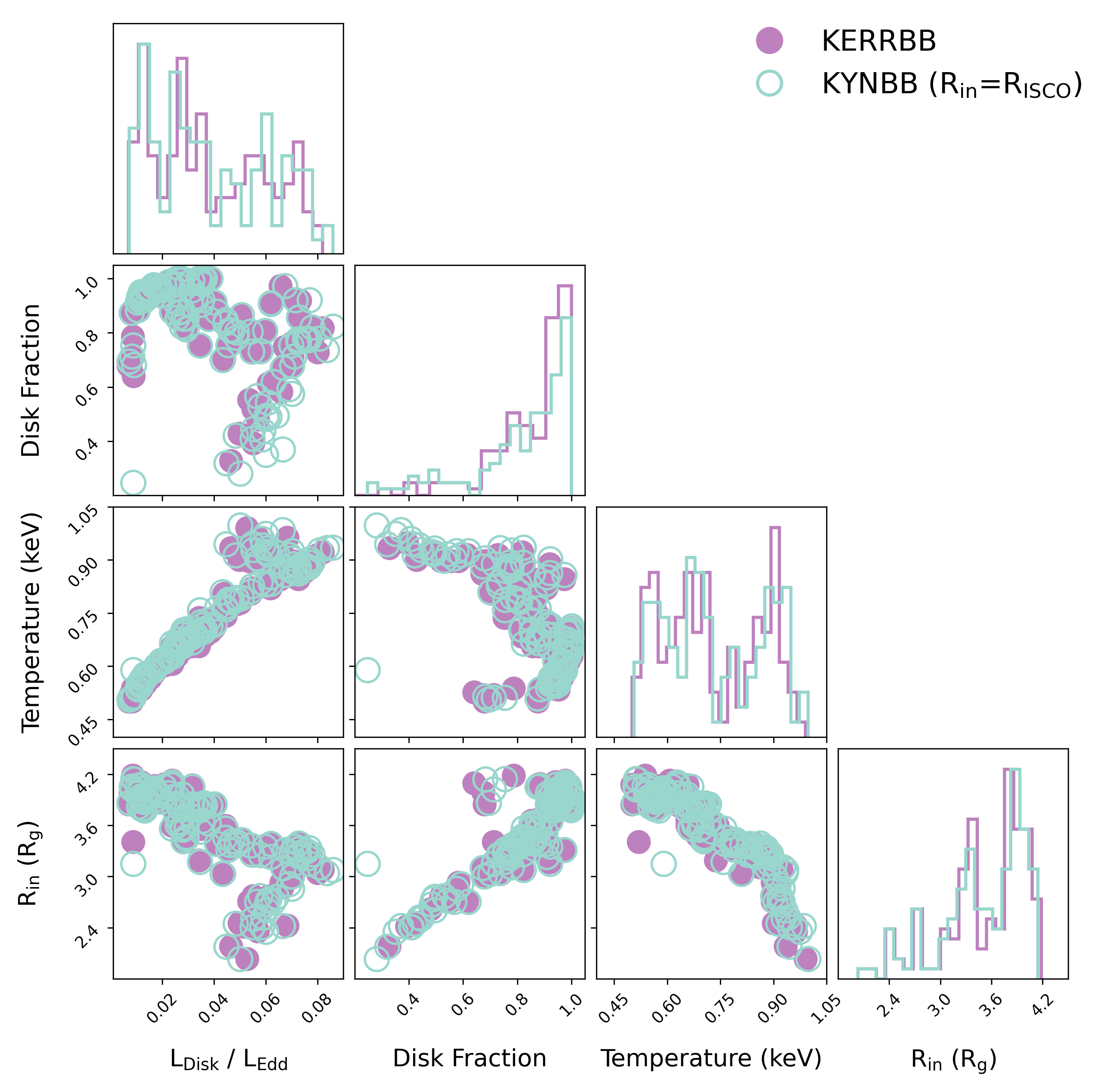}
    \caption{\textit{Top}: Corner plot presenting parameters (disk luminosity $\mathrm{L_{Disk}} / \mathrm{L_{Edd}}$, inner edge of the accretion disk $\rm R_{in}$, disk fraction (see Equation\ref{disk_fraction}) and disk temperature) for fits comparing \texttt{KERRBB} with \texttt{KYNBB} with $\rm R_{in} = \rm R_{ISCO} $.}
    \label{fig:corner-kynbb-kerrbb}
    \textit{} 
\end{figure*}

We present the comparison between the two models in FIGURE \ref{fig:corner-kynbb-kerrbb} and the evolution of other parameters of the system throughout the outburst. We first attempted to fix the spin at $\rm a_{*}=0.7$ following the results from the X-ray spectral continuum fitting method by \citet{Shafee06}. This assumed spin value resulted in significantly worse fit statistics for $\sim$ 89\% of observations which produced reduced $\rm \chi^{2}$ values much larger than 2.0. We tested how \texttt{KERRBB} behaved with different spin values and we did not observe the model favoring a specific value of spin over all possible values. These measured spin values were observed to span a parameter space between $\rm 0.52<a_{*}<0.94$ which corresponds to values that cannot be explained only by uncertainties introduced by the fitting procedure. This behavior in measured spin can be interpreted as a result of the notable evolution of the innermost edge of the accretion disk throughout the entire outburst of GRO J1655-40 (Yilmaz et al. 2022, in preparation).

We compared the two models using calculated disk fractions (Equation \ref{disk_fraction}), disk luminosity $\rm L_{Disk} / \rm L_{Edd}$, black hole spin and temperature values by setting the inner edge of the accretion disk parameter in \texttt{KYNBB} at $\rm R_{ISCO}$. We used the spin values measured by both \texttt{KERRBB} and \texttt{KYNBB} to show the evolution of $\rm R_{in}$ calculated using Equation \ref{r_isco} with respect to other parameters. The change in $\rm R_{in}$ with respect to the disk temperature has a clear trend indicating a decrease of $\rm R_{in}$ with increasing temperature values. This non-linear trend in the evolution of $\rm R_{in}$ was not observed with respect to $\mathrm{L_{Disk}} / \mathrm{L_{Edd}}$. There's a clear overturn in $\rm R_{in}$ values at $\rm R_{in}\sim 3 \; R_{g} $ and $\mathrm{L_{Disk}} / \mathrm{L_{Edd}} \sim \rm 0.08$, after which $\rm R_{in}$ starts to decrease with decreasing $\mathrm{L_{Disk}} / \mathrm{L_{Edd}}$. This behavior after the turning point contradicts the expected evolution of the inner edge of the accretion disk with respect to decreasing mass accretion rates where the disk is expected to be truncated at larger radii compared to higher mass accretion rates. This observed deviation in $\rm R_{in}$ is similar to the previously reported ${L_{\rm {Disk}}}$ - $T$ relations observed for many BHXRBs by \citet{Gier04, Done07, Dunn08,Dunn10, Dunn11}.

On average, both models produced consistent results for the black hole spin (4.5\% difference on average), disk temperature, disk fraction and disk luminosity for the same accretion rates assuming $\rm R_{in}=\rm R_{ISCO}$. There's an anti-correlation between the disk luminosity $\mathrm{L_{Disk}} / \mathrm{L_{Edd}}$ and the disk fraction. This anti-correlation contradicts the expected relation between these parameters. With increasing luminosity, the disk component is expected to start to dominate over the non-thermal component which results in increasing disk fraction (Equation \ref{disk_fraction}). Throughout the 2005 outburst, GRO J1655-40 was observed to enter an unusually soft or hyper-soft state \citep{Uttley15}. In this unexpected state, a strong accretion disc wind was detected with  Chandra X-ray HETG \citep{Miller08}. This sudden suppression of the non-thermal component could be explained by this Compton-thick accretion disk wind and it was even suggested that it might indicate intrinsic luminosities above the Eddington limit due to such high obscuration \citep{Neilsen16}. On the other hand, with increasing luminosity from the LHS, the source enters a state where the luminosity is very high but a steep-powerlaw starts to dominate over the thermal component resulting in an inverse relationship between the disk luminosity and the disk fractions calculated from the unabsorbed components of the total model.

\section{Conclusions}\label{conclusions}

Our analysis showed that both \texttt{KERRBB} and \texttt{KYNBB} provide identical results when $\mathrm{R_{in}}=\mathrm{R_{ISCO}}$ is assumed. Fitting results, shown in FIGURE \ref{fig:corner-kynbb-kerrbb}, show only small statistical fluctuations in some cases which can be explained by small differences in the reduced $\rm \chi^{2}$ values corresponding to different local minima in \texttt{XSPEC}.

Our analysis showed that a variable black hole spin parameter was necessary to obtain reasonable fit results for the entire 2005 outburst of GRO J1655-40. This behavior in the spin parameter could be interpreted as an actual change in the innermost edge of the accretion disk, possibly not even extending down to $\mathrm{R_{in}}$. While the $\mathrm{R_{in}}=\mathrm{R_{ISCO}}$ assumption in \texttt{KERRBB} has been widely adopted for the X-ray spectral continuum fitting method for over a decade, a variable inner edge of the disk might suggest that this standard assumption might not hold true throughout the entire set of different accretion states. While it is accepted that the inner edge of the accretion disk extends down to $\mathrm{R_{ISCO}}$ in the highest luminosity state, we still observed a notable variance in calculated $\mathrm{R_{in}}$ values. This variability in the innermost edge of the accretion disk, $\mathrm{R_{in}}$, will be further investigated in a follow-up work.

\backmatter

\section*{Acknowledgments}

A.Y., J.S., and P.G.B. acknowledge financial support from the \fundingAgency{Czech Science Foundation} under Project No. \fundingNumber{19-05599Y}. M.B. and M.D. thank for the support 
from the \fundingAgency{Czech Science Foundation} under Project No. \fundingNumber{21-06825X}.
A.Y., P.G.B., M.B., M.D., and J.S. also acknowledge the institutional support from \fundingAgency{Astronomical Institute of the Czech Academy of Sciences} with \fundingNumber{RVO:6798581}.

\bibliography{Wiley-ASNA}

\begin{thebibliography}{}

\bibitem [\protect \citeauthoryear {%
{Arnaud}%
}{%
{Arnaud}%
}{%
{\protect \APACyear {1996}}%
}]{%
Arnaud96}
\APACinsertmetastar {%
Arnaud96}%
\begin{APACrefauthors}%
{Arnaud}, K\BPBI A.%
\end{APACrefauthors}%
\unskip\
\newblock
\APACrefYearMonthDay{1996}{{\APACmonth{01}}}{},
\newblock
{\BBOQ}\APACrefatitle {{XSPEC: The First Ten Years}} {{XSPEC: The First Ten
  Years}}.{\BBCQ}
\newblock
\BIn{} G\BPBI H.~{Jacoby}\ \BBA {} J.~{Barnes}\ (\BEDS), \APACrefbtitle
  {Astronomical Data Analysis Software and Systems V} {Astronomical Data
  Analysis Software and Systems V}\ \BVOL~101, \BPG~17.
\PrintBackRefs{\CurrentBib}

\bibitem [\protect \citeauthoryear {%
{Beer}%
\ \BBA {} {Podsiadlowski}%
}{%
{Beer}%
\ \BBA {} {Podsiadlowski}%
}{%
{\protect \APACyear {2002}}%
}]{%
Beer02}
\APACinsertmetastar {%
Beer02}%
\begin{APACrefauthors}%
{Beer}, M\BPBI E.%
\BCBT {}\ \BBA {} {Podsiadlowski}, P.%
\end{APACrefauthors}%
\unskip\
\newblock
\APACrefYearMonthDay{2002}{{\APACmonth{03}}}{},
\newblock
\unskip
\newblock
\APACjournalVolNumPages{\mnras}{331}{2}{351-360}.
\newblock
\begin{APACrefDOI} \doi{10.1046/j.1365-8711.2002.05189.x} \end{APACrefDOI}
\PrintBackRefs{\CurrentBib}

\bibitem [\protect \citeauthoryear {%
{Belloni}%
, {Motta}%
\BCBL {}\ \BBA {} {Mu{\~n}oz-Darias}%
}{%
{Belloni}%
\ \protect \BOthers {.}}{%
{\protect \APACyear {2011}}%
}]{%
Belloni11}
\APACinsertmetastar {%
Belloni11}%
\begin{APACrefauthors}%
{Belloni}, T\BPBI M.%
, {Motta}, S\BPBI E.%
\BCBL {}\ \BBA {} {Mu{\~n}oz-Darias}, T.%
\end{APACrefauthors}%
\unskip\
\newblock
\APACrefYearMonthDay{2011}{{\APACmonth{09}}}{},
\newblock
\unskip
\newblock
\APACjournalVolNumPages{Bulletin of the Astronomical Society of
  India}{39}{3}{409-428}.
\PrintBackRefs{\CurrentBib}

\bibitem [\protect \citeauthoryear {%
{Done}%
, {Gierli{\'n}ski}%
\BCBL {}\ \BBA {} {Kubota}%
}{%
{Done}%
\ \protect \BOthers {.}}{%
{\protect \APACyear {2007}}%
}]{%
Done07}
\APACinsertmetastar {%
Done07}%
\begin{APACrefauthors}%
{Done}, C.%
, {Gierli{\'n}ski}, M.%
\BCBL {}\ \BBA {} {Kubota}, A.%
\end{APACrefauthors}%
\unskip\
\newblock
\APACrefYearMonthDay{2007}{{\APACmonth{12}}}{},
\newblock
\unskip
\newblock
\APACjournalVolNumPages{\aapr}{15}{1}{1-66}.
\newblock
\begin{APACrefDOI} \doi{10.1007/s00159-007-0006-1} \end{APACrefDOI}
\PrintBackRefs{\CurrentBib}

\bibitem [\protect \citeauthoryear {%
{Dov{\v{c}}iak}%
, {Karas}%
\BCBL {}\ \BBA {} {Yaqoob}%
}{%
{Dov{\v{c}}iak}%
\ \protect \BOthers {.}}{%
{\protect \APACyear {2004}}%
}]{%
Dovciak04}
\APACinsertmetastar {%
Dovciak04}%
\begin{APACrefauthors}%
{Dov{\v{c}}iak}, M.%
, {Karas}, V.%
\BCBL {}\ \BBA {} {Yaqoob}, T.%
\end{APACrefauthors}%
\unskip\
\newblock
\APACrefYearMonthDay{2004}{{\APACmonth{07}}}{},
\newblock
\unskip
\newblock
\APACjournalVolNumPages{\apjs}{153}{1}{205-221}.
\newblock
\begin{APACrefDOI} \doi{10.1086/421115} \end{APACrefDOI}
\PrintBackRefs{\CurrentBib}

\bibitem [\protect \citeauthoryear {%
{Dunn}%
, {Fender}%
, {K{\"o}rding}%
, {Belloni}%
\BCBL {}\ \BBA {} {Cabanac}%
}{%
{Dunn}%
\ \protect \BOthers {.}}{%
{\protect \APACyear {2010}}%
}]{%
Dunn10}
\APACinsertmetastar {%
Dunn10}%
\begin{APACrefauthors}%
{Dunn}, R\BPBI J\BPBI H.%
, {Fender}, R\BPBI P.%
, {K{\"o}rding}, E\BPBI G.%
, {Belloni}, T.%
\BCBL {}\ \BBA {} {Cabanac}, C.%
\end{APACrefauthors}%
\unskip\
\newblock
\APACrefYearMonthDay{2010}{{\APACmonth{03}}}{},
\newblock
\unskip
\newblock
\APACjournalVolNumPages{\mnras}{403}{1}{61-82}.
\newblock
\begin{APACrefDOI} \doi{10.1111/j.1365-2966.2010.16114.x} \end{APACrefDOI}
\PrintBackRefs{\CurrentBib}

\bibitem [\protect \citeauthoryear {%
{Dunn}%
, {Fender}%
, {K{\"o}rding}%
, {Belloni}%
\BCBL {}\ \BBA {} {Merloni}%
}{%
{Dunn}%
\ \protect \BOthers {.}}{%
{\protect \APACyear {2011}}%
}]{%
Dunn11}
\APACinsertmetastar {%
Dunn11}%
\begin{APACrefauthors}%
{Dunn}, R\BPBI J\BPBI H.%
, {Fender}, R\BPBI P.%
, {K{\"o}rding}, E\BPBI G.%
, {Belloni}, T.%
\BCBL {}\ \BBA {} {Merloni}, A.%
\end{APACrefauthors}%
\unskip\
\newblock
\APACrefYearMonthDay{2011}{{\APACmonth{02}}}{},
\newblock
\unskip
\newblock
\APACjournalVolNumPages{\mnras}{411}{1}{337-348}.
\newblock
\begin{APACrefDOI} \doi{10.1111/j.1365-2966.2010.17687.x} \end{APACrefDOI}
\PrintBackRefs{\CurrentBib}

\bibitem [\protect \citeauthoryear {%
{Dunn}%
, {Fender}%
, {K{\"o}rding}%
, {Cabanac}%
\BCBL {}\ \BBA {} {Belloni}%
}{%
{Dunn}%
\ \protect \BOthers {.}}{%
{\protect \APACyear {2008}}%
}]{%
Dunn08}
\APACinsertmetastar {%
Dunn08}%
\begin{APACrefauthors}%
{Dunn}, R\BPBI J\BPBI H.%
, {Fender}, R\BPBI P.%
, {K{\"o}rding}, E\BPBI G.%
, {Cabanac}, C.%
\BCBL {}\ \BBA {} {Belloni}, T.%
\end{APACrefauthors}%
\unskip\
\newblock
\APACrefYearMonthDay{2008}{{\APACmonth{06}}}{},
\newblock
\unskip
\newblock
\APACjournalVolNumPages{\mnras}{387}{2}{545-563}.
\newblock
\begin{APACrefDOI} \doi{10.1111/j.1365-2966.2008.13258.x} \end{APACrefDOI}
\PrintBackRefs{\CurrentBib}

\bibitem [\protect \citeauthoryear {%
R.~{Fender}%
\ \BBA {} {Belloni}%
}{%
R.~{Fender}%
\ \BBA {} {Belloni}%
}{%
{\protect \APACyear {2012}}%
}]{%
Fender12}
\APACinsertmetastar {%
Fender12}%
\begin{APACrefauthors}%
{Fender}, R.%
\BCBT {}\ \BBA {} {Belloni}, T.%
\end{APACrefauthors}%
\unskip\
\newblock
\APACrefYearMonthDay{2012}{{\APACmonth{08}}}{},
\newblock
\unskip
\newblock
\APACjournalVolNumPages{Science}{337}{6094}{540}.
\newblock
\begin{APACrefDOI} \doi{10.1126/science.1221790} \end{APACrefDOI}
\PrintBackRefs{\CurrentBib}

\bibitem [\protect \citeauthoryear {%
R.~{Fender}%
\ \BBA {} {Mu{\~n}oz-Darias}%
}{%
R.~{Fender}%
\ \BBA {} {Mu{\~n}oz-Darias}%
}{%
{\protect \APACyear {2016}}%
}]{%
Fender16}
\APACinsertmetastar {%
Fender16}%
\begin{APACrefauthors}%
{Fender}, R.%
\BCBT {}\ \BBA {} {Mu{\~n}oz-Darias}, T.%
\end{APACrefauthors}%
\unskip\
\newblock
\APACrefYearMonthDay{2016}{}{},
\newblock
{\BBOQ}\APACrefatitle {{The Balance of Power: Accretion and Feedback in Stellar
  Mass Black Holes}} {{The Balance of Power: Accretion and Feedback in Stellar
  Mass Black Holes}}.{\BBCQ}
\newblock
\BIn{} F.~{Haardt}, V.~{Gorini}, U.~{Moschella}\BCBL {}\ \BOthers {.}\ (\BEDS),
  \APACrefbtitle {Lecture Notes in Physics, Berlin Springer Verlag} {Lecture
  Notes in Physics, Berlin Springer Verlag}\ \BVOL~905, \BPG~65.
\newblock
\begin{APACrefDOI} \doi{10.1007/978-3-319-19416-5\_3} \end{APACrefDOI}
\PrintBackRefs{\CurrentBib}

\bibitem [\protect \citeauthoryear {%
R\BPBI P.~{Fender}%
, {Belloni}%
\BCBL {}\ \BBA {} {Gallo}%
}{%
R\BPBI P.~{Fender}%
\ \protect \BOthers {.}}{%
{\protect \APACyear {2004}}%
}]{%
Fender04}
\APACinsertmetastar {%
Fender04}%
\begin{APACrefauthors}%
{Fender}, R\BPBI P.%
, {Belloni}, T\BPBI M.%
\BCBL {}\ \BBA {} {Gallo}, E.%
\end{APACrefauthors}%
\unskip\
\newblock
\APACrefYearMonthDay{2004}{{\APACmonth{12}}}{},
\newblock
\unskip
\newblock
\APACjournalVolNumPages{\mnras}{355}{4}{1105-1118}.
\newblock
\begin{APACrefDOI} \doi{10.1111/j.1365-2966.2004.08384.x} \end{APACrefDOI}
\PrintBackRefs{\CurrentBib}

\bibitem [\protect \citeauthoryear {%
{Foellmi}%
, {Depagne}%
, {Dall}%
\BCBL {}\ \BBA {} {Mirabel}%
}{%
{Foellmi}%
\ \protect \BOthers {.}}{%
{\protect \APACyear {2006}}%
}]{%
Foellmi06}
\APACinsertmetastar {%
Foellmi06}%
\begin{APACrefauthors}%
{Foellmi}, C.%
, {Depagne}, E.%
, {Dall}, T\BPBI H.%
\BCBL {}\ \BBA {} {Mirabel}, I\BPBI F.%
\end{APACrefauthors}%
\unskip\
\newblock
\APACrefYearMonthDay{2006}{{\APACmonth{10}}}{},
\newblock
\unskip
\newblock
\APACjournalVolNumPages{\aap}{457}{1}{249-255}.
\newblock
\begin{APACrefDOI} \doi{10.1051/0004-6361:20054686} \end{APACrefDOI}
\PrintBackRefs{\CurrentBib}

\bibitem [\protect \citeauthoryear {%
{Garc{\'\i}a}%
, {McClintock}%
, {Steiner}%
, {Remillard}%
\BCBL {}\ \BBA {} {Grinberg}%
}{%
{Garc{\'\i}a}%
\ \protect \BOthers {.}}{%
{\protect \APACyear {2014}}%
}]{%
Garcia14}
\APACinsertmetastar {%
Garcia14}%
\begin{APACrefauthors}%
{Garc{\'\i}a}, J\BPBI A.%
, {McClintock}, J\BPBI E.%
, {Steiner}, J\BPBI F.%
, {Remillard}, R\BPBI A.%
\BCBL {}\ \BBA {} {Grinberg}, V.%
\end{APACrefauthors}%
\unskip\
\newblock
\APACrefYearMonthDay{2014}{{\APACmonth{10}}}{},
\newblock
\unskip
\newblock
\APACjournalVolNumPages{\apj}{794}{1}{73}.
\newblock
\begin{APACrefDOI} \doi{10.1088/0004-637X/794/1/73} \end{APACrefDOI}
\PrintBackRefs{\CurrentBib}

\bibitem [\protect \citeauthoryear {%
{Gierli{\'n}ski}%
\ \BBA {} {Done}%
}{%
{Gierli{\'n}ski}%
\ \BBA {} {Done}%
}{%
{\protect \APACyear {2004}}%
}]{%
Gier04}
\APACinsertmetastar {%
Gier04}%
\begin{APACrefauthors}%
{Gierli{\'n}ski}, M.%
\BCBT {}\ \BBA {} {Done}, C.%
\end{APACrefauthors}%
\unskip\
\newblock
\APACrefYearMonthDay{2004}{{\APACmonth{01}}}{},
\newblock
\unskip
\newblock
\APACjournalVolNumPages{\mnras}{347}{3}{885-894}.
\newblock
\begin{APACrefDOI} \doi{10.1111/j.1365-2966.2004.07266.x} \end{APACrefDOI}
\PrintBackRefs{\CurrentBib}

\bibitem [\protect \citeauthoryear {%
{Hjellming}%
\ \BBA {} {Rupen}%
}{%
{Hjellming}%
\ \BBA {} {Rupen}%
}{%
{\protect \APACyear {1995}}%
}]{%
Hjellming95}
\APACinsertmetastar {%
Hjellming95}%
\begin{APACrefauthors}%
{Hjellming}, R\BPBI M.%
\BCBT {}\ \BBA {} {Rupen}, M\BPBI P.%
\end{APACrefauthors}%
\unskip\
\newblock
\APACrefYearMonthDay{1995}{{\APACmonth{06}}}{},
\newblock
\unskip
\newblock
\APACjournalVolNumPages{\nat}{375}{6531}{464-468}.
\newblock
\begin{APACrefDOI} \doi{10.1038/375464a0} \end{APACrefDOI}
\PrintBackRefs{\CurrentBib}

\bibitem [\protect \citeauthoryear {%
{Homan}%
\ \BBA {} {Belloni}%
}{%
{Homan}%
\ \BBA {} {Belloni}%
}{%
{\protect \APACyear {2005}}%
}]{%
Homan05}
\APACinsertmetastar {%
Homan05}%
\begin{APACrefauthors}%
{Homan}, J.%
\BCBT {}\ \BBA {} {Belloni}, T.%
\end{APACrefauthors}%
\unskip\
\newblock
\APACrefYearMonthDay{2005}{{\APACmonth{11}}}{},
\newblock
\unskip
\newblock
\APACjournalVolNumPages{\apss}{300}{1-3}{107-117}.
\newblock
\begin{APACrefDOI} \doi{10.1007/s10509-005-1197-4} \end{APACrefDOI}
\PrintBackRefs{\CurrentBib}

\bibitem [\protect \citeauthoryear {%
{Jahoda}%
\ \protect \BOthers {.}}{%
{Jahoda}%
\ \protect \BOthers {.}}{%
{\protect \APACyear {2006}}%
}]{%
Jahoda06}
\APACinsertmetastar {%
Jahoda06}%
\begin{APACrefauthors}%
{Jahoda}, K.%
, {Markwardt}, C\BPBI B.%
, {Radeva}, Y.%
\ et al.\end{APACrefauthors}%
\unskip\
\newblock
\APACrefYearMonthDay{2006}{{\APACmonth{04}}}{},
\newblock
\unskip
\newblock
\APACjournalVolNumPages{\apjs}{163}{2}{401-423}.
\newblock
\begin{APACrefDOI} \doi{10.1086/500659} \end{APACrefDOI}
\PrintBackRefs{\CurrentBib}

\bibitem [\protect \citeauthoryear {%
{K{\"o}rding}%
, {Jester}%
\BCBL {}\ \BBA {} {Fender}%
}{%
{K{\"o}rding}%
\ \protect \BOthers {.}}{%
{\protect \APACyear {2006}}%
}]{%
Kording06}
\APACinsertmetastar {%
Kording06}%
\begin{APACrefauthors}%
{K{\"o}rding}, E\BPBI G.%
, {Jester}, S.%
\BCBL {}\ \BBA {} {Fender}, R.%
\end{APACrefauthors}%
\unskip\
\newblock
\APACrefYearMonthDay{2006}{{\APACmonth{11}}}{},
\newblock
\unskip
\newblock
\APACjournalVolNumPages{\mnras}{372}{3}{1366-1378}.
\newblock
\begin{APACrefDOI} \doi{10.1111/j.1365-2966.2006.10954.x} \end{APACrefDOI}
\PrintBackRefs{\CurrentBib}

\bibitem [\protect \citeauthoryear {%
{Kuulkers}%
\ \protect \BOthers {.}}{%
{Kuulkers}%
\ \protect \BOthers {.}}{%
{\protect \APACyear {2000}}%
}]{%
Kuulkers00}
\APACinsertmetastar {%
Kuulkers00}%
\begin{APACrefauthors}%
{Kuulkers}, E.%
, {in't Zand}, J\BPBI J\BPBI M.%
, {Cornelisse}, R.%
\ et al.\end{APACrefauthors}%
\unskip\
\newblock
\APACrefYearMonthDay{2000}{{\APACmonth{06}}}{},
\newblock
\unskip
\newblock
\APACjournalVolNumPages{\aap}{358}{}{993-1000}.
\PrintBackRefs{\CurrentBib}

\bibitem [\protect \citeauthoryear {%
{Li}%
, {Zimmerman}%
, {Narayan}%
\BCBL {}\ \BBA {} {McClintock}%
}{%
{Li}%
\ \protect \BOthers {.}}{%
{\protect \APACyear {2005}}%
}]{%
Li05}
\APACinsertmetastar {%
Li05}%
\begin{APACrefauthors}%
{Li}, L\BHBI X.%
, {Zimmerman}, E\BPBI R.%
, {Narayan}, R.%
\BCBL {}\ \BBA {} {McClintock}, J\BPBI E.%
\end{APACrefauthors}%
\unskip\
\newblock
\APACrefYearMonthDay{2005}{{\APACmonth{04}}}{},
\newblock
\unskip
\newblock
\APACjournalVolNumPages{\apjs}{157}{2}{335-370}.
\newblock
\begin{APACrefDOI} \doi{10.1086/428089} \end{APACrefDOI}
\PrintBackRefs{\CurrentBib}

\bibitem [\protect \citeauthoryear {%
{Maccarone}%
}{%
{Maccarone}%
}{%
{\protect \APACyear {2002}}%
}]{%
Maccarone02}
\APACinsertmetastar {%
Maccarone02}%
\begin{APACrefauthors}%
{Maccarone}, T\BPBI J.%
\end{APACrefauthors}%
\unskip\
\newblock
\APACrefYearMonthDay{2002}{{\APACmonth{11}}}{},
\newblock
\unskip
\newblock
\APACjournalVolNumPages{\mnras}{336}{4}{1371-1376}.
\newblock
\begin{APACrefDOI} \doi{10.1046/j.1365-8711.2002.05876.x} \end{APACrefDOI}
\PrintBackRefs{\CurrentBib}

\bibitem [\protect \citeauthoryear {%
{Markwardt}%
\ \BBA {} {Swank}%
}{%
{Markwardt}%
\ \BBA {} {Swank}%
}{%
{\protect \APACyear {2005}}%
}]{%
Markwardt05}
\APACinsertmetastar {%
Markwardt05}%
\begin{APACrefauthors}%
{Markwardt}, C\BPBI B.%
\BCBT {}\ \BBA {} {Swank}, J\BPBI H.%
\end{APACrefauthors}%
\unskip\
\newblock
\APACrefYearMonthDay{2005}{{\APACmonth{02}}}{},
\newblock
\unskip
\newblock
\APACjournalVolNumPages{The Astronomer's Telegram}{414}{}{1}.
\PrintBackRefs{\CurrentBib}

\bibitem [\protect \citeauthoryear {%
{Miller}%
\ \protect \BOthers {.}}{%
{Miller}%
\ \protect \BOthers {.}}{%
{\protect \APACyear {2008}}%
}]{%
Miller08}
\APACinsertmetastar {%
Miller08}%
\begin{APACrefauthors}%
{Miller}, J\BPBI M.%
, {Raymond}, J.%
, {Reynolds}, C\BPBI S.%
, {Fabian}, A\BPBI C.%
, {Kallman}, T\BPBI R.%
\BCBL {}\ \BBA {} {Homan}, J.%
\end{APACrefauthors}%
\unskip\
\newblock
\APACrefYearMonthDay{2008}{{\APACmonth{06}}}{},
\newblock
\unskip
\newblock
\APACjournalVolNumPages{\apj}{680}{2}{1359-1377}.
\newblock
\begin{APACrefDOI} \doi{10.1086/588521} \end{APACrefDOI}
\PrintBackRefs{\CurrentBib}

\bibitem [\protect \citeauthoryear {%
{Motta}%
, {Belloni}%
, {Stella}%
, {Mu{\~n}oz-Darias}%
\BCBL {}\ \BBA {} {Fender}%
}{%
{Motta}%
\ \protect \BOthers {.}}{%
{\protect \APACyear {2014}}%
}]{%
Motta14}
\APACinsertmetastar {%
Motta14}%
\begin{APACrefauthors}%
{Motta}, S\BPBI E.%
, {Belloni}, T\BPBI M.%
, {Stella}, L.%
, {Mu{\~n}oz-Darias}, T.%
\BCBL {}\ \BBA {} {Fender}, R.%
\end{APACrefauthors}%
\unskip\
\newblock
\APACrefYearMonthDay{2014}{{\APACmonth{01}}}{},
\newblock
\unskip
\newblock
\APACjournalVolNumPages{\mnras}{437}{3}{2554-2565}.
\newblock
\begin{APACrefDOI} \doi{10.1093/mnras/stt2068} \end{APACrefDOI}
\PrintBackRefs{\CurrentBib}

\bibitem [\protect \citeauthoryear {%
{Neilsen}%
, {Rahoui}%
, {Homan}%
\BCBL {}\ \BBA {} {Buxton}%
}{%
{Neilsen}%
\ \protect \BOthers {.}}{%
{\protect \APACyear {2016}}%
}]{%
Neilsen16}
\APACinsertmetastar {%
Neilsen16}%
\begin{APACrefauthors}%
{Neilsen}, J.%
, {Rahoui}, F.%
, {Homan}, J.%
\BCBL {}\ \BBA {} {Buxton}, M.%
\end{APACrefauthors}%
\unskip\
\newblock
\APACrefYearMonthDay{2016}{{\APACmonth{05}}}{},
\newblock
\unskip
\newblock
\APACjournalVolNumPages{\apj}{822}{1}{20}.
\newblock
\begin{APACrefDOI} \doi{10.3847/0004-637X/822/1/20} \end{APACrefDOI}
\PrintBackRefs{\CurrentBib}

\bibitem [\protect \citeauthoryear {%
{Novikov}%
\ \BBA {} {Thorne}%
}{%
{Novikov}%
\ \BBA {} {Thorne}%
}{%
{\protect \APACyear {1973}}%
}]{%
Novikov73}
\APACinsertmetastar {%
Novikov73}%
\begin{APACrefauthors}%
{Novikov}, I\BPBI D.%
\BCBT {}\ \BBA {} {Thorne}, K\BPBI S.%
\end{APACrefauthors}%
\unskip\
\newblock
\APACrefYearMonthDay{1973}{{\APACmonth{01}}}{},
\newblock
{\BBOQ}\APACrefatitle {{Astrophysics of black holes.}} {{Astrophysics of black
  holes.}}{\BBCQ}
\newblock
\BIn{} \APACrefbtitle {Black Holes (Les Astres Occlus)} {Black Holes (Les
  Astres Occlus)}\ \BPG~343-450.
\PrintBackRefs{\CurrentBib}

\bibitem [\protect \citeauthoryear {%
{Orosz}%
\ \BBA {} {Bailyn}%
}{%
{Orosz}%
\ \BBA {} {Bailyn}%
}{%
{\protect \APACyear {1997}}%
}]{%
Orosz97}
\APACinsertmetastar {%
Orosz97}%
\begin{APACrefauthors}%
{Orosz}, J\BPBI A.%
\BCBT {}\ \BBA {} {Bailyn}, C\BPBI D.%
\end{APACrefauthors}%
\unskip\
\newblock
\APACrefYearMonthDay{1997}{{\APACmonth{03}}}{},
\newblock
\unskip
\newblock
\APACjournalVolNumPages{\apj}{477}{2}{876-896}.
\newblock
\begin{APACrefDOI} \doi{10.1086/303741} \end{APACrefDOI}
\PrintBackRefs{\CurrentBib}

\bibitem [\protect \citeauthoryear {%
{Reis}%
, {Fabian}%
, {Ross}%
\BCBL {}\ \BBA {} {Miller}%
}{%
{Reis}%
\ \protect \BOthers {.}}{%
{\protect \APACyear {2009}}%
}]{%
Reis09}
\APACinsertmetastar {%
Reis09}%
\begin{APACrefauthors}%
{Reis}, R\BPBI C.%
, {Fabian}, A\BPBI C.%
, {Ross}, R\BPBI R.%
\BCBL {}\ \BBA {} {Miller}, J\BPBI M.%
\end{APACrefauthors}%
\unskip\
\newblock
\APACrefYearMonthDay{2009}{{\APACmonth{05}}}{},
\newblock
\unskip
\newblock
\APACjournalVolNumPages{\mnras}{395}{3}{1257-1264}.
\newblock
\begin{APACrefDOI} \doi{10.1111/j.1365-2966.2009.14622.x} \end{APACrefDOI}
\PrintBackRefs{\CurrentBib}

\bibitem [\protect \citeauthoryear {%
{Remillard}%
\ \BBA {} {McClintock}%
}{%
{Remillard}%
\ \BBA {} {McClintock}%
}{%
{\protect \APACyear {2006}}%
}]{%
Remillard06}
\APACinsertmetastar {%
Remillard06}%
\begin{APACrefauthors}%
{Remillard}, R\BPBI A.%
\BCBT {}\ \BBA {} {McClintock}, J\BPBI E.%
\end{APACrefauthors}%
\unskip\
\newblock
\APACrefYearMonthDay{2006}{{\APACmonth{09}}}{},
\newblock
\unskip
\newblock
\APACjournalVolNumPages{\araa}{44}{1}{49-92}.
\newblock
\begin{APACrefDOI} \doi{10.1146/annurev.astro.44.051905.092532}
  \end{APACrefDOI}
\PrintBackRefs{\CurrentBib}

\bibitem [\protect \citeauthoryear {%
{Shafee}%
\ \protect \BOthers {.}}{%
{Shafee}%
\ \protect \BOthers {.}}{%
{\protect \APACyear {2006}}%
}]{%
Shafee06}
\APACinsertmetastar {%
Shafee06}%
\begin{APACrefauthors}%
{Shafee}, R.%
, {McClintock}, J\BPBI E.%
, {Narayan}, R.%
, {Davis}, S\BPBI W.%
, {Li}, L\BHBI X.%
\BCBL {}\ \BBA {} {Remillard}, R\BPBI A.%
\end{APACrefauthors}%
\unskip\
\newblock
\APACrefYearMonthDay{2006}{{\APACmonth{01}}}{},
\newblock
\unskip
\newblock
\APACjournalVolNumPages{\apjl}{636}{2}{L113-L116}.
\newblock
\begin{APACrefDOI} \doi{10.1086/498938} \end{APACrefDOI}
\PrintBackRefs{\CurrentBib}

\bibitem [\protect \citeauthoryear {%
{Shakura}%
\ \BBA {} {Sunyaev}%
}{%
{Shakura}%
\ \BBA {} {Sunyaev}%
}{%
{\protect \APACyear {1973}}%
}]{%
Shakura73}
\APACinsertmetastar {%
Shakura73}%
\begin{APACrefauthors}%
{Shakura}, N\BPBI I.%
\BCBT {}\ \BBA {} {Sunyaev}, R\BPBI A.%
\end{APACrefauthors}%
\unskip\
\newblock
\APACrefYearMonthDay{1973}{}{},
\newblock
\unskip
\newblock
\APACjournalVolNumPages{\aap}{24}{}{337-355}.
\PrintBackRefs{\CurrentBib}

\bibitem [\protect \citeauthoryear {%
{Stella}%
\ \BBA {} {Vietri}%
}{%
{Stella}%
\ \BBA {} {Vietri}%
}{%
{\protect \APACyear {1998}}%
}]{%
Stella98}
\APACinsertmetastar {%
Stella98}%
\begin{APACrefauthors}%
{Stella}, L.%
\BCBT {}\ \BBA {} {Vietri}, M.%
\end{APACrefauthors}%
\unskip\
\newblock
\APACrefYearMonthDay{1998}{{\APACmonth{01}}}{},
\newblock
\unskip
\newblock
\APACjournalVolNumPages{\apjl}{492}{1}{L59-L62}.
\newblock
\begin{APACrefDOI} \doi{10.1086/311075} \end{APACrefDOI}
\PrintBackRefs{\CurrentBib}

\bibitem [\protect \citeauthoryear {%
{Stella}%
\ \BBA {} {Vietri}%
}{%
{Stella}%
\ \BBA {} {Vietri}%
}{%
{\protect \APACyear {1999}}%
}]{%
Stella99}
\APACinsertmetastar {%
Stella99}%
\begin{APACrefauthors}%
{Stella}, L.%
\BCBT {}\ \BBA {} {Vietri}, M.%
\end{APACrefauthors}%
\unskip\
\newblock
\APACrefYearMonthDay{1999}{{\APACmonth{01}}}{},
\newblock
\unskip
\newblock
\APACjournalVolNumPages{\prl}{82}{1}{17-20}.
\newblock
\begin{APACrefDOI} \doi{10.1103/PhysRevLett.82.17} \end{APACrefDOI}
\PrintBackRefs{\CurrentBib}

\bibitem [\protect \citeauthoryear {%
{Uttley}%
\ \BBA {} {Klein-Wolt}%
}{%
{Uttley}%
\ \BBA {} {Klein-Wolt}%
}{%
{\protect \APACyear {2015}}%
}]{%
Uttley15}
\APACinsertmetastar {%
Uttley15}%
\begin{APACrefauthors}%
{Uttley}, P.%
\BCBT {}\ \BBA {} {Klein-Wolt}, M.%
\end{APACrefauthors}%
\unskip\
\newblock
\APACrefYearMonthDay{2015}{{\APACmonth{07}}}{},
\newblock
\unskip
\newblock
\APACjournalVolNumPages{\mnras}{451}{1}{475-485}.
\newblock
\begin{APACrefDOI} \doi{10.1093/mnras/stv978} \end{APACrefDOI}
\PrintBackRefs{\CurrentBib}

\bibitem [\protect \citeauthoryear {%
{Zhang}%
\ \protect \BOthers {.}}{%
{Zhang}%
\ \protect \BOthers {.}}{%
{\protect \APACyear {1994}}%
}]{%
Zhang94}
\APACinsertmetastar {%
Zhang94}%
\begin{APACrefauthors}%
{Zhang}, S\BPBI N.%
, {Wilson}, C\BPBI A.%
, {Harmon}, B\BPBI A.%
\ et al.\end{APACrefauthors}%
\unskip\
\newblock
\APACrefYearMonthDay{1994}{{\APACmonth{08}}}{},
\newblock
\unskip
\newblock
\APACjournalVolNumPages{\iaucirc}{6046}{}{1}.
\PrintBackRefs{\CurrentBib}

\end{thebibliography}

\end{document}